\documentclass[conference]{IEEEtran}
\usepackage{graphicx}
\newtheorem{theorem}{Theorem}
\newtheorem{lemma}{Lemma}
\newtheorem{definition}{Definition}
\newtheorem{proposition}{Proposition}

\usepackage{float}
\usepackage{algorithmic}

\floatstyle{ruled}
\newfloat{algorithm}{tbp}{loa}
\floatname{algorithm}{Algorithm}
\IEEEoverridecommandlockouts
\title{Distributed Navigation Algorithms for Sensor Networks} 

\author{Chiranjeeb Buragohain, Divyakant Agrawal, Subhash Suri
\thanks{The research of the authors were supported by NSF grant
CCF-0514738 and Army Reasearch Organization grant DAAD19-03D0004.}
\\
Dept. of Computer Science, University of California, Santa Barbara, CA
93106, USA\\ \texttt{\{chiran,agrawal,suri\}@cs.ucsb.edu}}

\begin{document}

\maketitle 

\begin{abstract} We propose efficient distributed algorithms to aid
navigation of a user through a geographic area covered by sensors. The
sensors sense the level of danger at their locations and we use this
information to find a safe path for the user through the sensor
field. Traditional distributed navigation algorithms rely upon
flooding the whole network with packets to find an optimal safe
path. To reduce the communication expense, we introduce the concept of
a skeleton graph which is a sparse subset of the true sensor network
communication graph. Using skeleton graphs we show that it is possible
to find approximate safe paths with much lower communication cost. We
give tight theoretical guarantees on the quality of our approximation
and by simulation, show the effectiveness of our algorithms in
realistic sensor network situations.
\end{abstract}

\section{Introduction}
Recent advances in computing, communication, and related technologies
have resulted in significant interest in sensors and sensor
networks. Sensor networks are envisioned as a new link between the
physical world and the virtual world as it is modeled by computers,
networks, and information. In particular, once the physical world is
instrumented with sensors and sensor networks, the information-based
model of the physical world changes from a passive one to an active
one. Currently, however, sensors are primarily being deployed as
information collection points to monitor the physical environment.
But as the pace of innovation continues, in not too distant future, it
is likely that their scope will grow to allow interaction with as well
as control the physical world.

Most of the recent work in sensor networks has been confined to
developing technologies for monitoring the physical world.  Many
applications of sensor networks are indeed in this context: habitat
monitoring~\cite{szewczyk04analysis}, structural
monitoring~\cite{xu04wireless} and counter-sniper
systems~\cite{simon04sensor}.  Numerous research problems arise in the
context of such applications. In particular, a large body of recent
research activity in the area of sensor networks has focused on
various system level issues such as sensor
localization~\cite{moore04distributed}, medium access
protocols~\cite{polastre04low}, power-efficient
routing~\cite{aslam03three} and distributed query
processing~\cite{madden02tag}.

Only recently researchers have begun to explore more sophisticated
applications of sensor networks. Instead of viewing a sensor network
as a monitoring tool for the physical world, questions are being
explored if sensor and sensor networks can transition to become a
reactive system. For example, consider a world that has been
instrumented with sensors capable of detecting disruptive or
dangerous events (e.g., a chemical spill, a traffic accident)
and if and when such events occur, the system should be able to aid
navigation in the modified state of the physical world. Recently, Li et
al.~\cite{li03distributed} have proposed algorithms to answer exactly
this question: guiding the movements of a user through a sensor field in
the presence of dangers or obstacles. Their proposed solution
finds an optimal safest path, but it is based on the \emph{flooding} model
in which every sensor exchanges information with every other sensor.
This scheme does not scale well due to a very high communication cost.

In this paper, we propose more scalable solutions for the problem of
navigating a user in the presence of disruptions or hazards in a
sensor field. Our algorithms make two natural assumptions: (1) the
operational environment is assumed to have no large holes in the
coverage by sensors, and (2) an \emph{approximately optimal} safe path
is acceptable. Based on these two assumptions, we develop distributed
navigation algorithms that are very efficient in terms of their
communication cost; they find near-optimal paths with significantly
smaller communication (and, thus, energy) overhead.  The underlying
idea behind our scheme is to activate a sparse sub-network within the
dense sensor network and use this sparse network to solve the
navigation problem.  (We envision ``rotating'' the navigation duties
among the sensors so that a small fraction of the network is awake to
aid navigation at any point, while other sensors are in the sleep
mode.)
We explore two different ways to create
such sparse embeddings: the first one based on a uniform grid-like
mesh, and the other based on an adaptive mesh.

Our main result is that using sparse networks of size
$\mathcal{O}(n^{1/2+\epsilon})$, where $n$ is the total number of 
nodes in the networks, we can determine safe paths whose quality
(length, exposure, etc.) is within a small constant factor of the
optimal.

\section{Preliminaries and Related Work}
Let us assume that $n$ sensor \emph{nodes} are placed uniformly in a
square area.  We choose units of length such that the size of the area
is $n^{1/2}\times n^{1/2}$, i.e. on the average every unit area
contains one single sensor.  Every sensor can communicate with any
other sensor which is within radio range $r$ of it.  The number of
radio neighbors of a single sensor is not large, i.e. $ 1 < r \ll
n^{1/2}$.  Thus the sensor nodes form a logical graph with nodes as
vertices and communication links between neighbors as edges.  Also we
assume that each sensor knows its geographic location through some
localization algorithm.  The query for safe path is injected into the
system at a node which we shall call \emph{source}.  The query
specifies a destination coordinate.  The node closest to the
destination coordinate will be called \emph{destination}.  The safe
path is a path on the communication graph starting at the source and
ending at the destination.



\subsection{Metrics for Path Quality}



We consider two natural metrics for safe path: path length and
exposure.  We'll consider two examples to illustrate the relevance of
the two metrics in practice.  Suppose a dangerous chemical leak has
occurred in the region covered by the sensors.  We want the safe path
from source to destination be such that the maximum concentration of
the chemical on the path does not exceed a threshold $t$.  Thus we
define the \emph{danger zone} as the region where the chemical
concentration exceeds $t$.  The optimal path then is the shortest path
between the two points which stays outside the danger zone.  We call
this the \emph{shortest feasible path} (SFP).  In this paper, we shall
treat the hop distance in the network and true geometric distance
interchangeably.  Since our sensors are spread uniformly and they form
a dense network, such an assumption will not lead to gross
inaccuracies.

The next example is for a point like danger.  Let us assume that a
sensor detects the presence of an enemy soldier at some point in the
battlefield.  As we move through the battlefield, the enemy soldier
can detect us at a distance by some means, such as sight or sound, but
his capacity for detection goes down with distance.  Suppose the enemy
soldier is at the origin $(0,0)$ and he can detect us with probability
$\phi(x,y)$ if we are at the point $(x,y)$.  If we want to move from
the source to destination with the least probability of detection,
then we need to minimize the following quantity over all possible
paths $P$:
\begin{equation}
  \textrm{Probability of detection}  \propto S(P) \equiv \int_P \phi(x,y) d\ell.
\label{eqn-exp}
\end{equation}
We call the quantity $S(P)$ as the exposure for the path $P$ and the
optimal path as the \emph{minimum exposure path} (MEP).  To put it
more formally, the presence of enemy at $(0,0)$, creates a
\emph{potential} $\phi(x,y)$ at the point $(x,y)$ and we would like to
move along a path where the integrated potential along the path is
minimized.  The definition of the potential function $\phi(x,y)$
itself is arbitrary, but it should monotonically decrease as we move
away from the enemy position.  Assuming that the enemy is at the
origin, a convenient potential function is
\begin{equation}
\phi(x,y) = \frac{1}{(x^2+y^2)^{\beta/2}} \equiv \frac{1}{R^\beta},
\;\;\beta > 0, 
\label{eqn-pot}
\end{equation}
where $R$ is the Euclidean distance from the point of danger to the
point $(x,y)$.  For our purposes in this paper, we shall impose the
condition $\beta > 1$.
Another desirable property for the potential is the
\emph{superposition} property defined as follows.  If there are $k$
enemy points denoted by $1, 2, \ldots k$, then the total potential at
$(x,y)$ is defined as
\begin{equation}
  \phi_\mathrm{total}(x,y) = \sum_{i=1}^k \phi(x-x_i,y-y_i),
\label{eqn-pot-superposition}
\end{equation}
where $\phi(x-x_i,y-y_i)$ is the potential at $(x,y)$ due to enemy
point $i$ located at $(x_i, y_i)$.  



There are some important constraints that one needs to impose on the
complexity of the danger zone.  In general, if the side-length of the sensor
field is $\mathcal{O}(n^{1/2})$, one expects the length of any
shortest path be bounded by $\mathcal{O}(n^{1/2})$.  But one can
easily conjure up pathologically shaped danger zones for which the
length of the optimal path can be as long as $\mathcal{O}(n)$.  We
exclude such exceptional cases by imposing the constraint that the
perimeter of the danger zone be ``well behaved'' in the following
sense.  Let us consider a curve and a square box of size $x$ which
intersects the curve.  The well behavedness property restricts the
length of the curve inside that box.
\begin{definition}
\label{def-well-behaved}
A curve is well behaved, if for any square box of side $x$ that
intersects the curve, the length of the curve inside the box is less
than $cx$ for some constant $c > 1$, and for all $x$.\footnote{This
condition is same as saying that the curve has fractal dimension 1.}
\end{definition}
This is not a very stringent condition and any polygon of low
complexity satisfies it.  This property will be key in proving the
efficiency of our algorithms.  We also demand that the number of
distinct dangerous entities be a constant much smaller than $n^{1/2}$.

In summary, the problem which we shall address in this paper is as
follows: given an area covered by sensors where one or more danger
zones exist, can we efficiently compute approximate shortest paths and
minimum exposure paths between any two points?

\subsection{The Skeleton Graph}
\begin{figure}
  \begin{center} \includegraphics[width=0.35\textwidth]{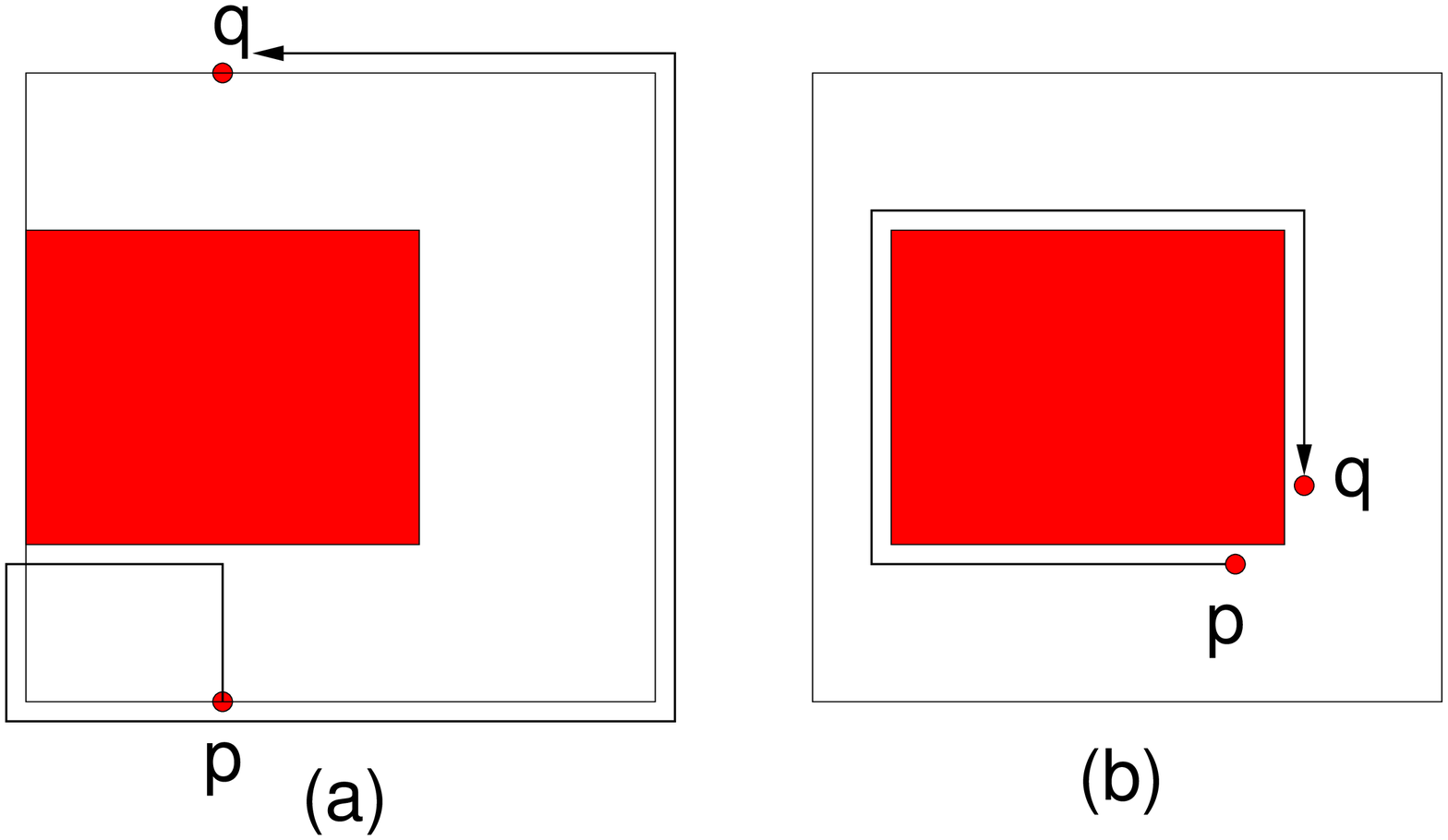}
  \end{center}
\caption{Bad performance of greedy geographic routing schemes: the
source is $p$ and destination is $q$, while the shaded area represents
the danger zone.  In (a), the principle of perimeter traversal leads
to a traversal of the whole field.  In (b), the length of the
path found is as large as the perimeter of the danger zone.}
\label{fig-gpsr}
\end{figure}

Navigating a sensor field in the presence of danger zones is a problem
which is similar to path planning in the presence of obstacles.  There
are two obvious ways one can approach this problem: a greedy
geographic scheme similar to GPSR routing \cite{karp00gpsr} and
exhaustive search.  In a geographic scheme, one would greedily move
towards the destination and traverse around the danger zones
encountered on the way.  This scheme has very low communication
overhead, but can lead to highly suboptimal paths as shown in
Fig. \ref{fig-gpsr}.
The global exhaustive search algorithm floods the network with packets
to carry out a Breadth-First-Search (BFS) on the communication graph.
Obviously this algorithm is optimal in terms of path length, but very
expensive in terms of communication cost.


Our solution in this direction splits up the problem into two pieces.
The first step is to construct a reduced graph with fewer nodes from
the full communication graph.  We call this smaller graph the
\emph{skeleton graph}.  The second part is to carry out a search on
the skeleton graph to find shortest paths and minimal exposure paths
\emph{over the skeleton graph only}.  If the skeleton graph is small
in size, then even carrying out an exhaustive search over the skeleton
graph will not be very expensive in terms of communication.
The requirements that we impose on the skeleton graph are as follows:
(i) If a safe path exists in the original graph, a safe path exists in
the skeleton graph too. (ii) The quality of the path found in the
skeleton graph is comparable to the optimal path.



Our main contribution in this work lies in constructing a small sized
skeleton graph from the main communication graph.  Once the skeleton
graph is constructed, the problem of finding the optimal paths on
these graphs can be achieved with a set of simple algorithms.  These
algorithms are \emph{reactive} algorithms rather than \emph{proactive}
algorithms.  In other words, we do not maintain path information in the
system; only when a query is made, path discovery takes place.  We
briefly discuss these algorithms below.  These algorithms are
applicable to \emph{any} graph and not special to skeleton graphs in
any way.

\subsection{Shortest  Path Algorithm}
\label{sec-sfp-algo}
This algorithm is nothing but BFS over the communication graph.  The
graph is flooded with search packets starting from the source.  Every
packet contains two fields which specify how many hops it has traveled
from the source and the last node visited.  When a node receives a
search packet, it increments the hop count by 1 and forwards the
packet to the other neighbors.  Every node maintains a distance variable
which counts the minimum number of hops to the source and a parent
pointer which points to the node via which the minimum hop search
packet was received.  If a node receives multiple search packets from
the source, only packets with smallest hop counts are forwarded.  When
BFS terminates, every node knows its distance to the source and its
parent pointer points to its parent along the path towards the source.

Note that in a shortest path computation by BFS, the number of packets
transmitted by each node is exactly 1.  The first search packet that
arrives at a node is the packet which has traveled by the least
number of hops.  The packets which arrive later arrive by traveling
larger number of hops and hence are discarded.  This leads us to the
following proposition which bounds the communication cost of shortest
path discovery by BFS search.

\begin{proposition}
  \label{prop-bfs-cost}
In a network of $n$ nodes, the number of total packet transmissions
required for the shortest path algorithm is $\mathcal{O}(n)$.
\end{proposition}

\subsection{Minimum Exposure Path Algorithm}
\label{sec-mep-algo}

\begin{algorithm}
  \caption{\textsc{Minimum-Exposure}}
  \label{algo-min-exp}
  \begin{algorithmic}[1]
    \WHILE{TRUE}	    

    \STATE Receive($pkt$) from $neighbor$
    \STATE $pkt.exposure \gets pkt.exposure + self.potential$
    \IF{$pkt.exposure < minexposure$}
    \STATE $minexposure \gets pkt.exposure$
    \STATE $parent \gets neighbor$
    \STATE schedule $pkt$ for forwarding
    \ELSE
    \STATE Drop($pkt$)
    \ENDIF
    \IF {there is a scheduled packet}
    \STATE Transmit($pkt$)
    \ENDIF
    \ENDWHILE
  \end{algorithmic}
\end{algorithm}

Exposure computation relies upon the computation of potential
$\phi(x,y)$ first.  
We give a simple algorithm for potential calculation below.  Assume
that the potentials at each point are known.  Then minimum exposure
path calculation is very similar to the shortest path BFS algorithm.
In this case the path length is the total exposure, not total hop
count.  Search packets are injected into the network by the source and
nodes forward these packets to their neighbors.  Every search packet
carries with itself a variable \emph{exposure} which is just the sum
of potentials of the nodes it has passed through.  Thus any packet
contains within it the total exposure of the path that it has
traveled.  Just like BFS above, every node maintains a total exposure
field (\emph{minexposure}) and a parent pointer.  The variable
\emph{minexposure} measures the total exposure of the minimum exposure
path from the source.  Any packet which arrives at a node with total
exposure more than the value \emph{minexposure} at that node is not
forwarded.  Otherwise, the node updates the \emph{minexposure}
variable and forwards the packet.  The pseudocode is shown in
algorithm~\ref{algo-min-exp}.  When the algorithm terminates, every
node knows the exposure of the minimum exposure path to the source.

Now we give a simple algorithm to calculate the potential due to a
single danger point.  Potential due to multiple danger points can be
computed using the principle of superposition
(eqn. \ref{eqn-pot-superposition}).  Consider a sensor which detects
danger at its location.  This sensor floods the network with packets
for a BFS much like the shortest path calculation.  Thus every node on
the network learns its distance from the danger point and hence can
now compute the potential according to eqns. \ref{eqn-pot} and
\ref{eqn-pot-superposition}.

\subsection{Related Work}
Navigation and path planning has a long history as a robotics
\cite{latombe92robot} and computational geometry
\cite{berg00computational} problem.  The challenge for sensor network
environment is that path planning must be done in a distributed
manner.  The problem of route finding in ad-hoc networks is similar to
the problem that we address here.  Greedy Perimeter Stateless Routing
(GPSR) \cite{karp00gpsr} is a greedy routing strategy for ad hoc
networks which utilizes geographic information to find its
destination.  We have already discussed the unsuitability of
geographic schemes for navigation.  The alternative protocols like
AODV \cite{perkins00ad} and DSR \cite{johnson00dynamic} do not utilize
geographic information and instead flood the network with query
packets for finding routes.  Obviously such a flooding scheme is not
efficient for sensor networks.

The concept of minimum exposure path were introduced by Meguerdichian
et. al.~\cite{meguerdichian01exposure}.  Veltri
et. al.~\cite{veltri03minimal} has given heuristics to distributedly
compute minimal and maximal exposure paths in sensor networks.  Path
planning in the context of sensor networks was addressed by Li
et. al.~\cite{li03distributed} where they consider the problem of
finding minimum exposure path.  Their approach involves exhaustive
search over the whole network to find the minimal exposure path.
Recently Liu et.al.~\cite{liu04combs} have used the concept of
searching a sparse subgraph to implement algorithms for resource
discovery in sensor networks.  This work, which was carried out
independently of us, however doesn't address the problem of path
finding when parts of the sensor network is blocked due to danger.
Some of our work is inspired by the mesh generation problem
\cite{berg00computational,bern92mesh} in computational geometry.  


\section{Navigation Using Uniform Skeleton Graph}
\label{sec-uniform-skeleton-graph}

\begin{figure}
  \begin{center}
    \begin{tabular}{cc}
  \includegraphics[width=0.15\textwidth]{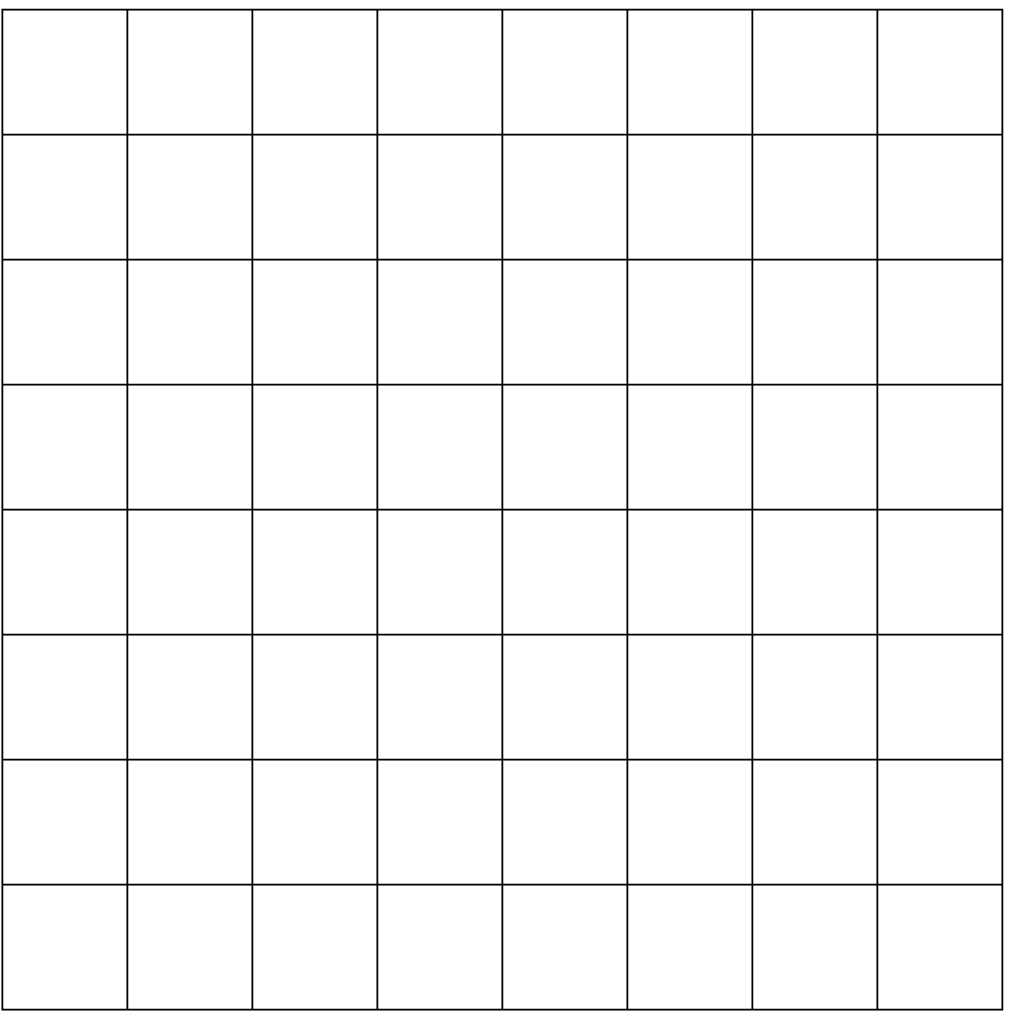} &      
  \includegraphics[width=0.25\textwidth]{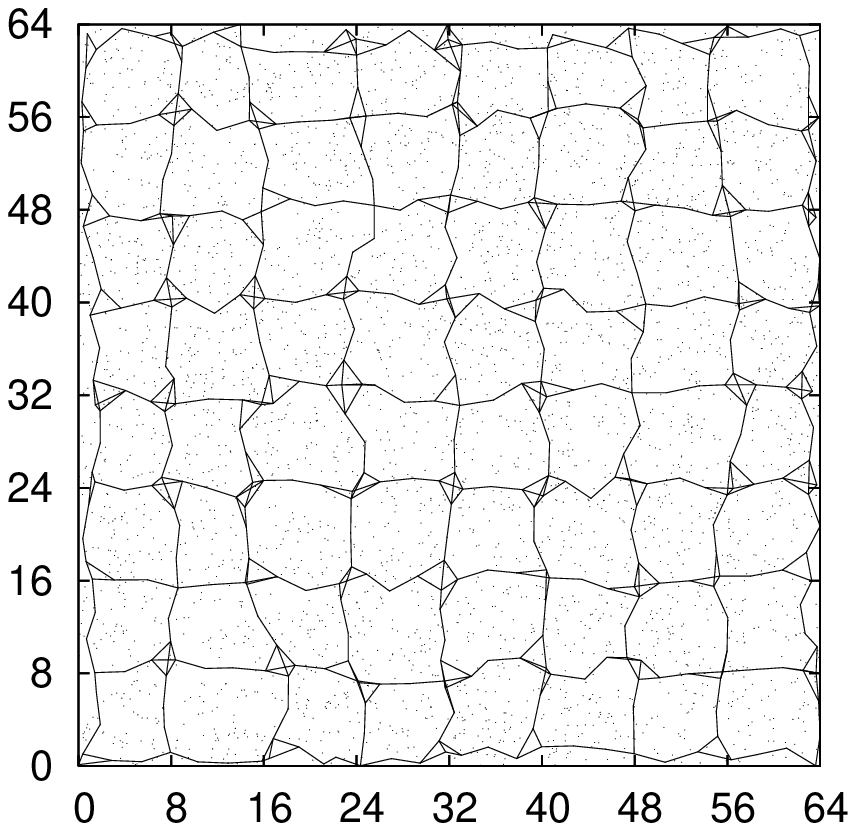}
    \end{tabular}
  \end{center}
\caption{A street map and the corresponding skeleton graph.  There are
4096 nodes with range 3 each.  The skeleton graph contains only 450
nodes.}
\label{fig-embed}
\end{figure}

Suppose we have an area $A$ covered with sensors.  Then the structure
of the skeleton graph can be explained most intuitively in terms of a
set of line segments inside this area $A$.  We call these segments
\emph{streets} and the collection of all the streets within the area a
\emph{street map}.  The nodes which are geographically ``close'' to
the streets constitute the skeleton graph.  We shall soon define this
idea of closeness more rigorously, but meanwhile an example will
clarify the concept.
Fig. \ref{fig-embed} shows a set of streets and the corresponding
skeleton graph.  All nodes which are not part of the skeleton graph
are put to sleep and they do not communicate with other nodes.  Thus
the skeleton graph is a small set of nodes which geographically span
the area $A$ and form within themselves a connected communication
graph.  The street map is an ideal geometric representation of the
skeleton graph and its communication links.  The exact algorithm for
\emph{embedding} a set of streets in a true network graph will be given in
Section \ref{sec-embedding}.  For theoretical purposes in this paper, we
shall use the skeleton graph and its abstract street map
representation interchangeably.

It is clear that given an arbitrary distribution of sensors and an
arbitrary street map, it is not possible to successfully embed the
street map in the communication graph.  In most realistic settings,
the sensors will be deployed in a random fashion, leading to an
expected-case uniform coverage of the field. Mathematically, we assume
that each sensor's location $(x,y)$ is a pair of independent random
variables distributed uniformly.  In fact, the only technical
requirement of our scheme is that the sensor field not have any large
holes in its coverage.  Under these conditions and a reasonable radio
communication range, it is possible to carry out the embedding.  We
shall address this issue is a more quantitative fashion in
sec. \ref{sec-expt}.


\begin{figure}
  \begin{center}
  \includegraphics[width=0.35\textwidth]{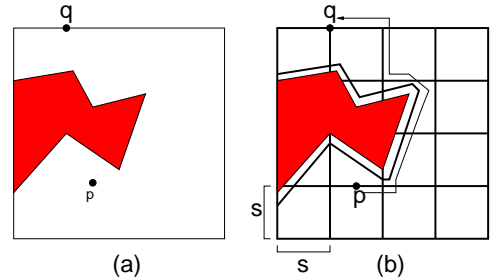}
  \end{center}
\caption{The street map for uniform skeleton graph.  In (a), we show
the danger zone as a shaded area and two points $p$ and $q$ between
which we seek a shortest path.  In (b) we see the street map with grid
size $s$ and the streets outlined in bold.  The shortest path between
$p$ and $q$ is is also shown.}
\label{fig-grid-map}
\end{figure}

In this section we introduce the uniform skeleton graph which contains
two classes of streets: \emph{grid streets} and \emph{perimeter
streets}.  The grid streets are a square grid of lines separated by
distance $s$ from each other.  An additional set of streets which
follow the perimeter of the danger zone is also included in the street
map and they are the perimeter streets.  Fig. \ref{fig-grid-map} shows
a single danger zone and the street map that results from it.

\subsection{The Uniform Skeleton Graph: Streets and Embeddings}
\label{sec-embedding}
We assume that all nodes know the value of $s$ which is the separation
between streets.  Then the embedding of the grid streets is achieved
as follows.  Let us imagine every grid street to be a strip of width
$w$ instead of being a line.  Since the nodes know their positions,
they can independently decide if they are within distance $w$ of any
grid street.  All nodes which lie on the strip include themselves in
the skeleton graph, while the other nodes go to sleep.  As long as $wr
> 1$, with high probability, the number of nodes lying along the
streets is enough to ensure that all the nodes lying along the streets
form a connected set.  The embedding of the streets can be optimized
by an additional step.  In general, the strip of width $w$ will
contain some redundant nodes which can be put to sleep without losing
connectivity.  To do this we assign the two nodes at each end of the
street to be source and destination.  The source carries out a BFS
search for the shortest path within this street to the destination.
Only the nodes which are on the shortest path from source to
destination are included in the skeleton graph.  

Note that a protocol like GPSR can be also used construct the grid
streets.  Let us assume that some node initiates the street
construction protocol.  Then using GPSR, we can send out street
construction packets along the perpendicular grid lines starting with
the initiating node.  All nodes which are touched by the construction
packets include themselves in the skeleton graph.  This method has
very low overhead for constructing skeleton graph, but it might
produce sub-optimal graphs in the presence of holes.

Next we turn to the embedding of the perimeter streets.  To do this,
the nodes which are on the danger zone boundary need to detect first
that they are on the boundary.  This is an easy problem to solve: if a
node realizes that it is in the danger zone, but it has at least one
neighbor outside the danger zone, then that node declares itself to be
at the boundary.  The nodes inside the danger zone can go to sleep.
The boundary nodes broadcast a ``wake-up'' message with lifetime of
$w$ hops to its neighbors.  Any node within $w$ hop of a boundary will
be awakened and added to the skeleton graph.  These nodes constitute
the perimeter streets.  Nodes inside the danger zone are always
excluded from the skeleton graph.

\begin{figure}
  \begin{center}
  \includegraphics[width=0.2\textwidth]{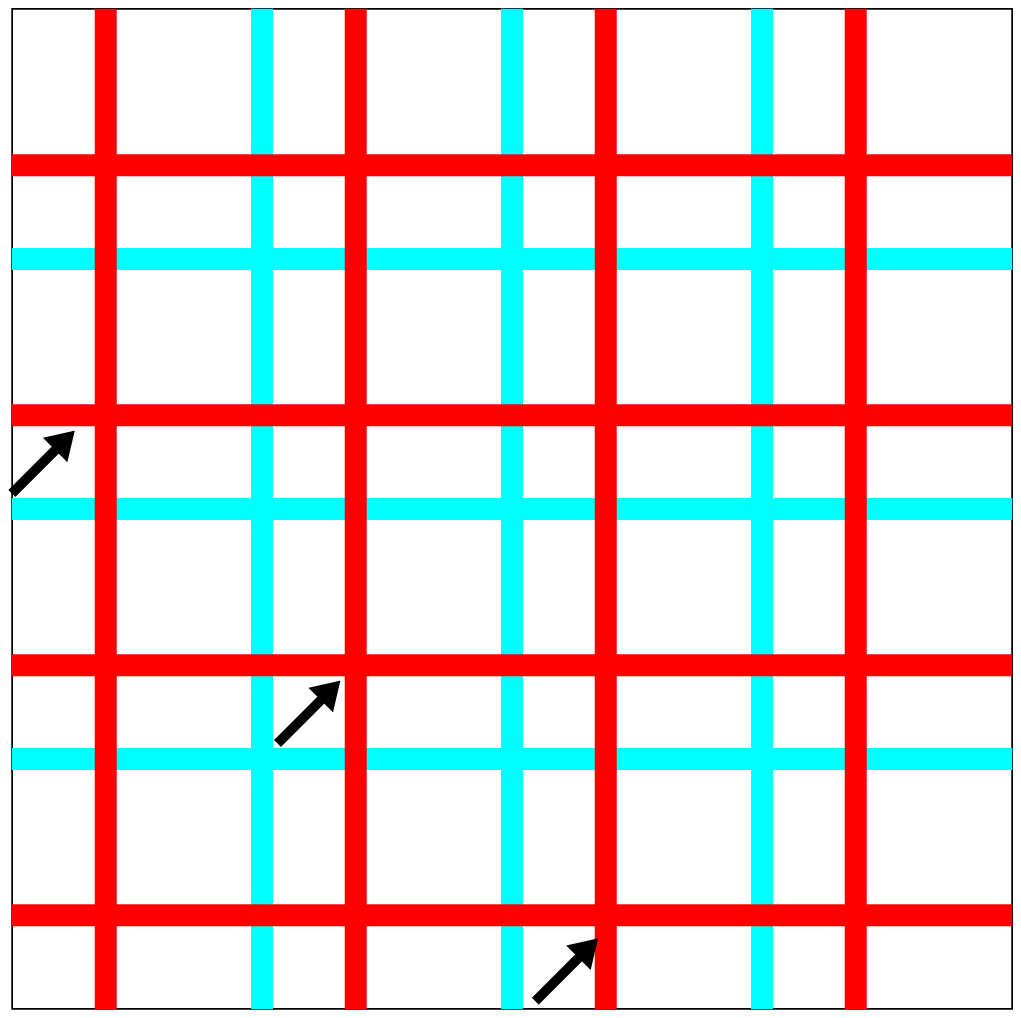}    
  \end{center}
\caption{Load balancing by shifting.}
\label{fig-shift}
\end{figure}

Once we have constructed the skeleton graph, the shortest path and the
minimum exposure paths can be constructed using the algorithms
described in sections \ref{sec-sfp-algo} and \ref{sec-mep-algo}.  Note
that although the skeleton graph requires only a small subset of the
nodes to participate in path finding, over time this set of nodes
might run out of energy prematurely compared to other nodes which are
not included in the skeleton graph.  This can be avoided by varying
the value of $s$, the street separation; or by shifting all the
streets by a constant amount in the diagonal direction as shown in
Fig. \ref{fig-shift}.  

\subsection{Path Discovery for Points not on Streets}

So far we have restricted our attention to finding shortest paths
between source and destination pairs which are on the streets.  What
can one do for source and destination pairs which do not lie on any
street?  There are two solutions.  The first solution is for the
source to initiate flooding to discover the closest street to it and
from then on, follow the streets for route discovery to the
destination.  If the destination does not lie on any street, then it
is enclosed in a square enclosed by four streets.  As soon as any
packet realizes that it is on the boundary of the square, then the
destination can be found by flooding that square.  This flooding adds
some overhead to path discovery, but this overhead is comparatively
low for long paths.

The second solution is to construct the streets on-demand rather than
to pre-compute them.  As soon as a source needs to discover a route to
the destination, it initiates construction of streets centered around
itself.  We can use GPSR to construct the grid streets as before.  The
benefit of this approach is that in this case, load balancing is
automatic because every path discovery query produces its own set of
streets.  As mentioned before, this method might be suboptimal if
there are significant holes in the communication graph.

\subsection{The Uniform Skeleton Graph: Basic Properties}
In this section we focus on the performance characteristics of these
algorithms and prove the approximation bounds.  We first prove the
following theorem which limits the size of the uniform skeleton graph,
and hence limits the total communication cost of a search in that
graph.

\begin{theorem}
The communication cost of discovering the shortest path in the uniform
skeleton graph is $\mathcal{O}(n^{1/2+\epsilon})$, for any $\epsilon$
such that $0 < \epsilon < 1/2$.
\end{theorem}
\begin{proof}
There are two sets of streets: the grid streets and the perimeter
streets.  Every grid street is of length $n^{1/2}$ and the number of
grid streets is $2\times n^{1/2}/s$.  Thus the total length of grid
streets is $\mathcal{O}(n/s)$.  Since the perimeter of a danger zone
is well behaved, the total length of the perimeter as well as the
perimeter streets is $\mathcal{O}(n^{1/2})$.  The width of the streets
$w$ is a constant of order unity, while $ 1 < s < n^{1/2}$.  Clearly,
the total street length is dominated by the grid streets.  Hence we
set $s = n^{1/2-\epsilon}$ and find that the total number of nodes in
the skeleton graph is $\mathcal{O}(n^{1/2+\epsilon}$).  Applying
proposition \ref{prop-bfs-cost}, we immediately see that the
communication cost must also be bound by
$\mathcal{O}(n^{1/2+\epsilon}$).
\end{proof}
Note that since $1 < s < n^{1/2}$, $\epsilon$ is constrained to lie
between 0 and 1/2.  The exact choice of $\epsilon$ involves a
trade-off between skeleton graph size and the quality of path found.
A larger value of $\epsilon$, gives better coverage of the area with
streets at the expense of involving large number of nodes in the path
search.

We now consider the quality of the approximate path length in the
uniform skeleton graph.  Let us first introduce some notation.  Given
any two points which are located on streets, there is an optimal path:
$P_\textrm{OPT}$ and a path along the streets which we call
$P_\textrm{USG}$.  Their lengths are $\ell_\textrm{OPT}$ and
$\ell_\textrm{USG}$ respectively.  The following theorem gives the
worst case bound on the length of $\ell_\mathrm{USG}$.
\begin{figure}
  \begin{center}
  \includegraphics[width=0.3\textwidth]{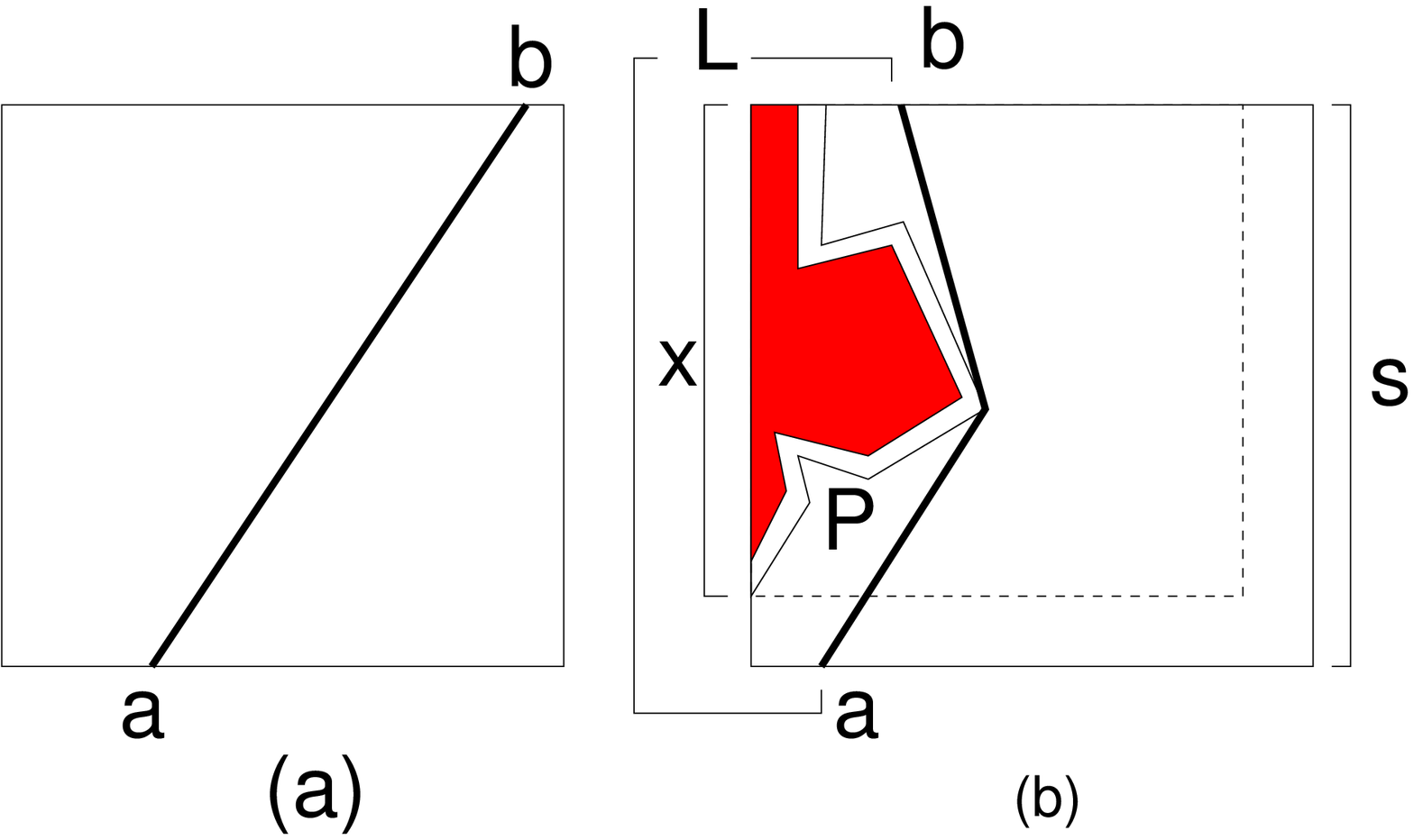}    
  \end{center}
\caption{Shortest path in uniform grid.}
\label{fig-grid-path}
\end{figure}

\begin{theorem}
For a path joining any two points located on the streets in uniform
skeleton graph,
\begin{displaymath}
\ell_\textrm{USG}/\ell_\textrm{OPT} \leq 2(1+c),
\end{displaymath}
where $c$ is the constant appearing in the definition of well
behavedness (Def. \ref{def-well-behaved}).
\label{thm-grid-path}
\end{theorem}
\begin{proof}
The optimal path goes through a sequence of grid street squares.  We
shall decompose the optimal path into segments, each of which is
contained completely within its own square.  The sides of the square
are the streets.  If we prove the bound on each square separately,
then the total path will also obey the required bound.  Consider a
segment of the optimal path that goes through a square on the grid and
it crosses the square at points $a$ and $b$
(Fig. \ref{fig-grid-path}).  Since the communication is restricted to
move along the streets, there are two paths to get from $a$ to $b$.
There are two cases to take care of while bounding the length of path
along the streets.

Fig. \ref{fig-grid-path} (a) exhibits the case when the boundary of
the square is free of danger.  In that case the shortest path from $a$
to $b$ along the streets is at most twice as long as the optimal path.

Fig. \ref{fig-grid-path} (b) shows the case when one of the sides of
the square is blocked by danger.  Let's say if the danger was not
there, then there would be a path of length $L$ from $a$ to $b$ along
the left side of the square.  Because of the danger on the edge, the
path is forced to traverse the perimeter street $P$ and hence becomes
longer.  We bound the length of the perimeter street as follows: if
the danger zone can be bound within a square of side $x$, then the
length of the perimeter street length is $cx$ by the well behaved
property.  So total length of the path is at most the sum of the
perimeter path length $cx$ and the street length $L$, i.e.
$\ell_\mathrm{USG} \leq cx + L$.  If $x< L$, $\ell_\textrm{OPT} \geq
L/2$ and then
\begin{displaymath}
\ell_\mathrm{USG} \leq cx + L \leq (c+1)L \leq 2(1+c) \ell_\textrm{OPT}.
\end{displaymath}
if $x > L$, then the optimal path length $\ell_\textrm{OPT} \geq x$ and
\begin{displaymath}
\ell_\mathrm{USG} \leq cx + L \leq (c+1)x \leq (1+c) \ell_\textrm{OPT}.
\end{displaymath}
The case when both sides of the square intersect the perimeter streets
can be handled in a similar fashion.
\end{proof}

\subsection{The Uniform Skeleton Graph: Exposure}

We now consider the minimum exposure path problem.
Let us denote the exposure along the true minimum exposure path be
$S_\mathrm{OPT}$.  The minimum exposure using only the skeleton graph
is $S_\mathrm{USG}$.  Before we explore the relation between optimal
and approximate exposure, let us prove a useful lemma.

\begin{figure}
  \begin{center}
  \includegraphics[width=0.25\textwidth]{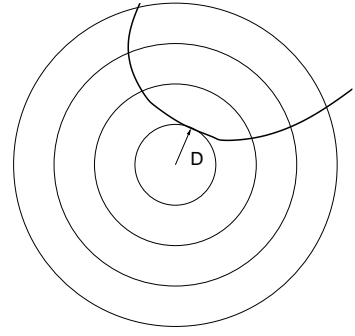}    
  \end{center}
\caption{Exposure along a path.}
\label{fig-exp}
\end{figure}

In this lemma, we shall consider a point danger and a path of length
$L$ which approaches at its closest to within distance $D$ of the
danger point (see Fig. \ref{fig-exp}).  The lemma gives an estimate of
the total exposure of this path.  Intuitively we can motivate this
lemma as follows: for $\beta > 1$, the potential dies fast as one goes
away from the danger.  So if the closest distance that the path
approaches the danger is $D$, then the major contribution to the
exposure comes from a region of size $D$ nearest to the danger.  The
contribution of the path outside this region contributes to the total
exposure only by a constant factor.
\begin{lemma}
For a well behaved path of length $L$ with minimum approach distance
$D$, and $\beta > 1$, the exposure of the path is given by
\begin{equation}
S  = \left\{
\begin{array}{ll}
c_1\frac{1}{D^{\beta-1}} &  L \geq D,\\
c_2\frac{L}{D^{\beta}} & L < D, 
\end{array}
\right.
\end{equation}
where $c_1$ and $c_2$ are constants.
\label{lemma-exp}
\end{lemma}
\begin{proof}
Consider a curve of length $L\geq D$ as shown in Fig. \ref{fig-exp}.
We divide the curve into segments by concentric circles of radius $D,
2D, 3D\ldots$  Since the curve is well behaved, the total length of
the curve \emph{inside} a circle of radius $kD$ is bounded by $c_rkD$
for some constant $c_r$.  Consider a segment of the curve contained
between circles of radius $kD$ and $(k+1)D$.  By the well behaved
assumption, the length of this segment is at most $c_aD$ for some
constant $c_a$.  So the exposure of this segment is bounded by
$\frac{c_aD}{(kD)^\beta}$.  The total exposure then is
\begin{equation}
S \leq \sum_{k=1}^\infty\frac{c_aD}{(kD)^\beta} =
\frac{c_a}{D^{\beta-1}}\sum_{k=1}^\infty\frac{1}{k^\beta}
\end{equation}
The sum on the RHS converges to a constant when $\beta > 1$.  The case
for $L < D$ is simple.  The potential is
$\frac{1}{D^\beta}$ and the length of the path is $L$.
The exposure immediately follows from that.
\end{proof}
\begin{figure}
  \begin{center}
  \includegraphics[width=0.25\textwidth]{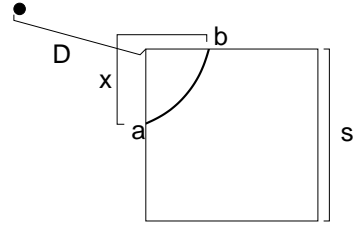}    
  \end{center}
\caption{Exposure along a path in the grid.}
\label{fig-grid-exp}
\end{figure}
The following theorem bounds the exposure performance of the uniform
skeleton graph scheme.
\begin{theorem}
For a path joining any two points located on the streets,
\begin{displaymath}
S_\mathrm{USG}/S_\mathrm{OPT} = \mathrm{const}.  
\end{displaymath}
\label{thm-grid-exp}
\end{theorem}
\begin{proof}
For the sake of brevity, here we give only an outline of the proof of
this theorem.  As in theorem \ref{thm-grid-path}, we shall decompose
the optimal path into segments wholly contained within a single square
and prove the bound for a single square.  For simplicity we assume
that there is a single point of danger as shown in
Fig. \ref{fig-grid-exp}.  Let the optimal exposure path cross the
square at points $a$ and $b$.  To go from $a$ to $b$ there are two
possible paths: a shorter path with exposure $S_1$ and a longer one
with exposure $S_2$.  Their respective lengths are $x$ and $4s-x$
where $s$ is the size of the square.  In terms of exposure, the short
path has the disadvantage of traversing a region of high potential,
while the longer path has the disadvantage of being long.


To compute $S_\textrm{USG}$, we consider the case $x > D$ first as
shown in Fig. \ref{fig-grid-exp}.  By lemma \ref{lemma-exp} the
exposures are as follows:
\begin{eqnarray}
  S_1 & = & \mathcal{O}\left(\frac{1}{D^{\beta-1}}\right), \; 
  S_2 =  \mathcal{O}\left(\frac{1}{(D+x)^{\beta-1}}\right)\\ 
  S_\textrm{OPT} &  = & \mathcal{O}\left(\frac{1}{(D+x)^{\beta-1}}\right).
\end{eqnarray}
The worst case results when $S_1 = S_2$, which implies that $x=
\mathcal{O}(D)$, i.e. $S_1, S_2$ and $S_\textrm{OPT}$ are within
constant factor of each other.   For $x < D$, 
\begin{eqnarray}
  S_1 & = & \mathcal{O}\left(\frac{x}{D^\beta}\right), \; 
  S_2 =  \mathcal{O}\left(\frac{1}{(D+x)^{\beta-1}}\right)\\ 
  S_\textrm{OPT} &  = & \mathcal{O}\left(\frac{x}{(D+x)^{\beta}}\right).
\end{eqnarray}
The worst case exposure results when $S_1 = S_2$, which implies that $x=
\mathcal{O}(D)$, i.e. $S_1, S_2$ and $S_\textrm{OPT}$ are within
constant factor of each other.

The case of multiple danger points is a simple generalization.  By the
principle of superposition, (eqn. \ref{eqn-pot-superposition}) the
total exposure of a path due to multiple danger points is equal to the
sum of exposures due to each danger point taken separately.  Thus the
proof above remains valid for multiple points of danger as well.
\end{proof}

\section{Navigation  Using Adaptive Skeleton Graph}
\label{sec-adaptive-skeleton-graph}
The uniform skeleton graph is simple and effective, but it is possible
to improve upon it.  The uniform skeleton graph puts streets with
uniform density (all streets have separation $s$) in every region,
without regard for the region's distance from the danger zones.
Intuitively, if a user wants to navigate an area with danger zones, it
will be useful if near the danger zones, the streets are placed close
together, while far away the street layout is much coarser.  Readers
familiar with computational geometry literature will recognize a
similarity of this problem to the problem of adaptive mesh generation
\cite{bern92mesh}.

\begin{figure}
  \begin{center}
  \includegraphics[width=0.3\textwidth]{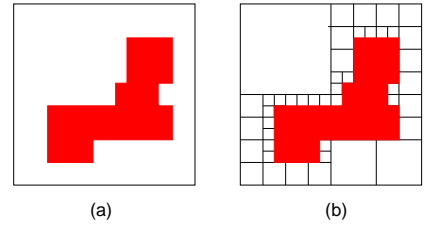}    
  \end{center}
\caption{Street map for a danger zone using a four level quadtree. (a)
shows the danger zone as the shaded area.  In (b) we see the quadtree
division so that boundary of the danger zone is completely aligned
with the quadtree.}
\label{fig-quad}
\end{figure}

\begin{figure}
  \begin{center}
  \includegraphics[width=0.2\textwidth]{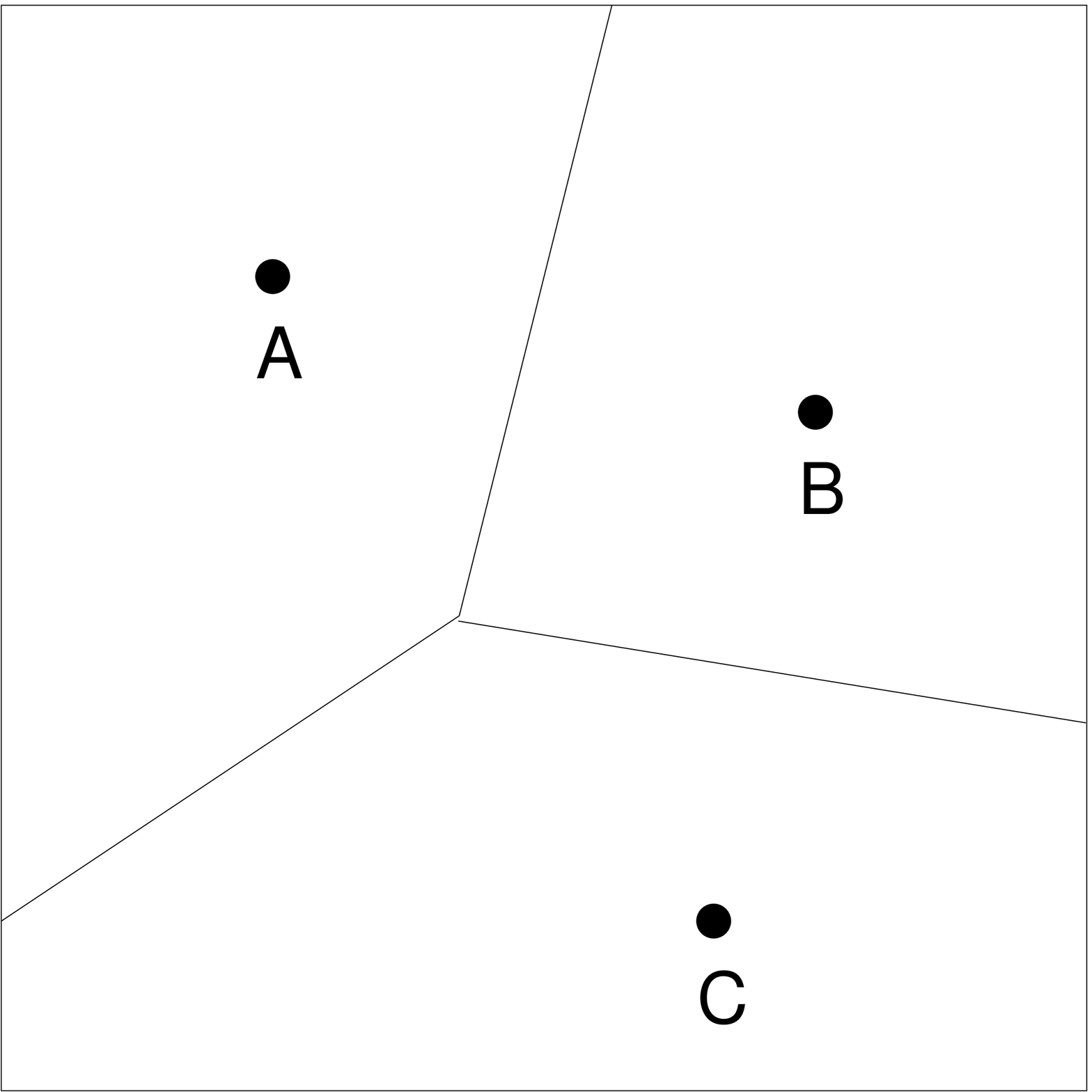}    
  \end{center}
\caption{Voronoi streets for three points of danger.}
\label{fig-voronoi}
\end{figure}

Let us now see how such a non-uniform adaptive street-map can be
produced.  For this discussion we assume that the danger zone boundary
is axis aligned.  Our street map will consist of a set of line
segments of length $1, 2, 4, \ldots, n^{1/2}$ which are also axis
aligned.  The logical representation of the street map can be best
done in terms of a quadtree.  This process is very similar to quadtree
mesh generation.  The whole $n^{1/2}\times n^{1/2}$ square area
corresponds to the root node in the quadtree.  We recursively divide
the area into quadtree cells until there is no quadtree cell whose
boundary is intersected by danger zone boundary.  The process is
illustrated in Fig. \ref{fig-quad}.  In the next section we show that
this adaptive construction is not only more efficient in terms of
number of nodes involved, but also has better guarantees on total path
length compared to the uniform skeleton graph.

Note that, having more detailed street map near the danger zones is
efficient for computing shortest paths, but it is not efficient for
computing minimum exposure paths.  Consider for example the three
points of danger inside the coverage area as shown in
Fig. \ref{fig-voronoi}.  Since the minimum exposure path should stay
as far away from the danger points as possible, intuitively the best
one can do is to move along the Voronoi edges
~\cite{berg00computational} for the three points.  Thus for computing
minimum exposure paths, we would want to compute Voronoi edges for the
danger points and embed them using the quadtree.

\subsection{The Adaptive Skeleton Graph: Streets and Embeddings}
The quadtree street map can be created in a distributed fashion, but
it requires a little coordination among sensor nodes.  At this point
we shall be a little loose with terminology and use the word node to
denote a true physical sensor as well as a node on the abstract
quadtree representation of the street map.  No confusion should arise
though, because the meaning will be clear from the context.

A quadtree node of level $k$ in physical terms consists of a square of
side length $2^k$.  The sensor nodes which lie within distance $w/2$
of the boundary of the square correspond to a \emph{cluster}.  We
shall assign a single node in the cluster to be a \emph{cluster
leader} for that cluster.  The cluster leader can be elected using any
suitable leader election algorithm.  Note that a single segment of an
edge in the quadtree can be part of several squares of different
sizes.  Thus a single sensor can belong to multiple clusters.  The
communication primitive required by an ordinary node is very simple.
It needs to be able to send a message to its cluster leader and
forward any message to its neighbor.  The cluster leader has more
responsibility.  It can communicate with its cluster by sending a
message which traverses the boundary of the square.  A cluster leader
also needs to know the leaders of its parent cluster and children
clusters

The quadtree is built recursively.  At the beginning the quadtree
consists of all the leaf squares and hence the skeleton graph consists
of all nodes.  If a cluster leader of a leaf square determines that
none of its nodes are within the danger zone, then it sends a message
informing its parent cluster leader of this fact.  If a parent cluster
determines that all its children are danger free, then it can instruct
its children to go to sleep.  This process repeats recursively up the
quadtree all the way to the root.  The skeleton graph then consists of
all the clusters which are still awake.

The embedding of Voronoi edges can be done in a very similar manner.
To compute nodes which are on the Voronoi edge, we adopt the following
algorithm.  Recall that every sensor located at a danger point carries
out a potential computation (Section \ref{sec-mep-algo}) which is
nothing but a BFS distance computation.  A node which finds that it is
equidistant from any two danger points declares itself to be on a
Voronoi edge.  Once the Voronoi edges are computed, embedding them
using a quadtree can be done as outlined above.

\subsection{The Adaptive Skeleton Graph: Properties}
The following theorem shows that the adaptive skeleton graph is highly
efficient in terms of total number of nodes in the graph.
\begin{theorem}
The communication cost of discovering the shortest path in the
adaptive skeleton graph is $\mathcal{O}(n^{1/2}\log n)$.
\end{theorem}
\begin{proof}
For simplicity we shall assume that there is only one danger zone with
perimeter length $p$.  By the assumption that the perimeter is well
behaved, $p = \mathcal{O}(n^{1/2})$.  Let us number the quadtree
levels as $0, 1, 2, \ldots $ with level 0 as the leaf level.  Thus the
quadtree level of $k$ corresponds to a square of side $2^k$.  Since
the perimeter length is $p$, the perimeter is adjacent to $p$ squares
of level 0.  By the same logic, the perimeter crosses $p/2$ nodes of
level 2, $p/4$ nodes of level 2 and so on.  In general the perimeter
crosses $p/2^k$ nodes of level $k$ (size $2^k$), and hence requires
$p/2^k$ nodes in its representation.  These $p/2^k$ nodes contribute a
total of $\mathcal{O}(p/2^k\times 2^k) = \mathcal{O}(p)$ length of
streets to the street map.  Since there are a total of $\log(n)$
levels in the quadtree, the total length of streets in the quadtree is
$\mathcal{O}(p\log n) = \mathcal{O}(n^{1/2}\log n)$.  The length of
streets in the quadtree immediately gives us the upper bound on the
number of nodes in the adaptive skeleton graph and by proposition
\ref{prop-bfs-cost}, the upper bound on the communication cost of the
shortest path computation.
\end{proof}

Now we turn to the issue of path lengths and exposure in the adaptive
skeleton graph.  Here we can mostly take over the discussion that we
have gone over in Section \ref{sec-uniform-skeleton-graph} and simplify
the proofs.  The following theorem shows that the adaptive skeleton
graph is very efficient in terms of shortest path.  Let the length for
the shortest path in the quadtree grid be $\ell_\textrm{ASG}$.
\begin{theorem}
For a path joining any two points located on the streets in the
adaptive skeleton graph, 
\begin{displaymath}
\ell_\textrm{ASG}/\ell_\textrm{OPT} \leq 2 .    
\end{displaymath}
\end{theorem}
\begin{proof}
The optimal path passes through a set of quadtree squares.  Unlike the
uniform skeleton graph, in the adaptive skeleton graph none of the
squares are intersected by the danger zone boundary.  Now consider a
segment of the optimal path going through a single square as shown in
Fig. \ref{fig-grid-path} (a).  It is clear that a path which sticks to
the sides of the square is at most twice long as the optimal path.
\end{proof}
The performance bound for the minimum exposure path for the adaptive
skeleton graph is identical to the uniform skeleton graph.  Note that
in theorem \ref{thm-grid-exp}, the size of the square itself did not
appear anywhere.  Hence that proof can serve without any modification
for the following theorem:
\begin{theorem}
For a path joining any two points located on the streets in adaptive
skeleton graph,
\begin{displaymath}
\frac{S_\mathrm{ASG}}{S_\mathrm{OPT}} = \mathrm{const},
\end{displaymath}
where $S_\mathrm{ASG}$ is the exposure for the adaptive skeleton graph.
\end{theorem}

\begin{figure}
  \begin{center}
  \includegraphics[width=0.3\textwidth]{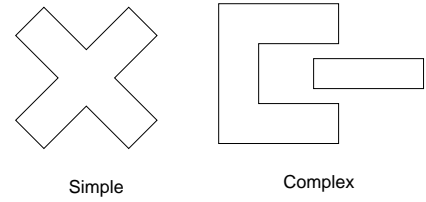}    
  \end{center}
\caption{Simple and complex danger zone shapes used to test shortest
path algorithms.}
\label{fig-obstacle}
\end{figure}

\section{Experimental Results}
\label{sec-expt}

We simulated our algorithms on simulated communication graph
topologies.  The simulation parameters are as follows.  We place $n$
sensor nodes in a $\sqrt{n}\times \sqrt{n}$ area.  The node
coordinates are random variables uniformly distributed within this
area.  The experiments were done with $n=1024, 4096$ and 16384 nodes.
Note that the average separation between nodes in our experiments is
1.  So the radio range decides the number of communication neighbors
of each node and is an indirect measure of node deployment density.
Experimentally we find that unless radio range is larger than 1.5, the
resulting graph is almost always disconnected.  Even when the
communication graph is connected, because of random fluctuations in
node density there are always large voids in the communication graph.
These voids are known to cause significant problems for geographic
routing protocols \cite{karp00gpsr}.  For our experiments we assume
that the radio range is 3 and in this range the occurrence of large
voids is rare.  Note that this is not a very dense deployment of
nodes.  For MICAz motes manufactured by the Crossbow
Corp. \cite{crossbow04micaz} which have radio range of 300ft, this
works out to a deployment where average node separation is 100ft.  

The
shortest path algorithms were implemented for two different types of
danger zones which we label \emph{simple} and \emph{complex}.  Their
shapes are shown in Fig. \ref{fig-obstacle}.
\begin{figure}
  \begin{center}
  \includegraphics[width=0.4\textwidth]{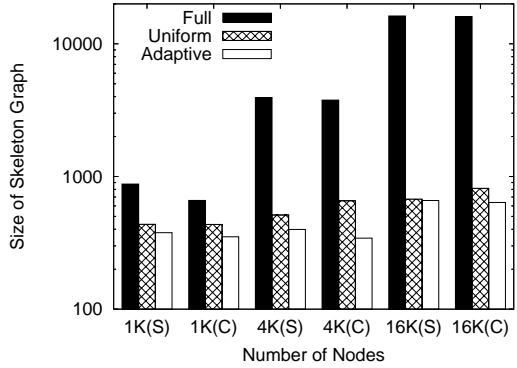}    
  \end{center}
  \caption{Number of nodes in skeleton graph for different sized
  networks.  The label 1K(S) means 1024 nodes with simple danger zone,
  16K(C) means 16384 nodes with complex danger zone etc.  The exponent
  $\epsilon = 0.05$.  Note that the vertical scale is logarithmic.}
  \label{fig-numnodes}
\end{figure}
The skeleton graph is a complex experimental system where one can
measure many relevant quantities such as the effect of varying size
and shapes of danger zones, effect of street separation $s$ on path
lengths and others.  In the interest of space, we only report the
results of a limited set of experiments which evaluate the size of
skeleton graphs and their performance in finding shortest paths and
minimum exposure paths.  
\subsection{Skeleton Graph Size}
In Fig. \ref{fig-numnodes}, we exhibit the size of the skeleton graph
for different shapes of danger zones and different number of sensors.
As we can see, the size of the skeleton graphs are much much smaller
than the full graph and this difference is more pronounced for larger
network sizes.  For a network of 1024 sensors and a simple danger
zone, the size of the adaptive skeleton graph is only 377, i.e. 37\%
of the original graph.  When we increase network size to 16384, there
are 659 nodes are in the adaptive skeleton graph ----which is only 4\%
of the full graph.  The uniform skeleton graph is slightly larger than
the adaptive graph, but this difference is not highly significant.
\begin{figure}
  \begin{center}
  \includegraphics[width=0.4\textwidth]{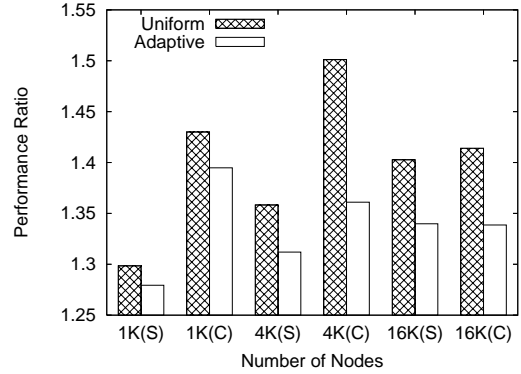}
  \end{center}
  \caption{Path length performance ratio of the uniform and adaptive
  skeleton graphs for different network sizes.  The label 1K(S) means
  1024 nodes with simple danger zone, 16K(C) means 16384 nodes with
  complex danger zone etc.  The exponent $\epsilon = 0.05$. }
  \label{fig-pathlen}
\end{figure}

\subsection{Shortest Path}
To evaluate the quality of the shortest path found in the skeleton
graph we generated a set of 200 random point pairs lying within the
sensor coverage area.  Let us assume that the lengths of the optimal
path and approximate skeleton graph paths are $\ell_\mathrm{OPT}$ and
$\ell_\mathrm{SG}$ respectively.  Then the efficiency of the algorithms
is defined by the $ \textrm{Path Length Performance Ratio} \equiv
\ell_\mathrm{SG}/\ell_\mathrm{OPT}$.
Closer the performance ratio to 1, better the algorithm.  In
Fig. \ref{fig-pathlen} we plot the average performance ratio for both
uniform and adaptive skeleton graphs.  The optimal path was found by
carrying out BFS over the full graph.  We see that for a large range
of network sizes and a combination of simple as well as complex danger
zones, the approximate path lengths are no worse than 50\% of the
optimal.  The adaptive skeleton graph performs better as expected.

\subsection{Minimum Exposure Path}
\begin{figure}
  \begin{center}
  \includegraphics[width=0.4\textwidth]{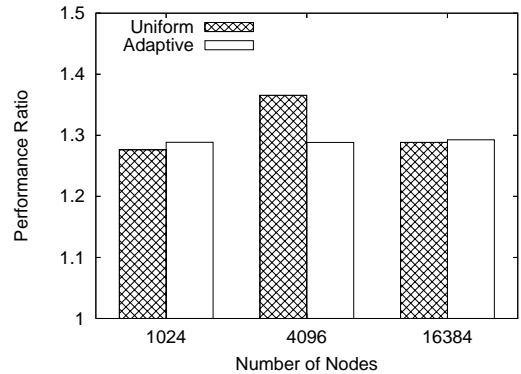}
  \end{center}
  \caption{Exposure performance ratio of the uniform and adaptive
  skeleton graphs for different network sizes.  The exponent $\epsilon
  = 0.2$. }
  \label{fig-pot}
\end{figure}
For minimum exposure path we generated 20 different scenarios, each of
which consists of three points of danger randomly placed inside the
coverage area.  For each set of three points we computed 10 minimum
exposure paths using the skeleton graph.  The exponent $\epsilon$ was
chosen such that the size of the uniform skeleton graph was roughly
equal to the size of the adaptive graph.  The optimal exposure was
calculated by BFS over the full graph as before.  If the exposures for
optimal and approximate paths are $S_\mathrm{OPT}$ and
$S_\mathrm{SG}$, we define the performance ratio as before to be $
\textrm{Exposure Performance Ratio} \equiv
S_\mathrm{SG}/S_\mathrm{OPT}$.  The average performance ratio is
plotted in Fig. \ref{fig-pot}.  As we can see both the uniform and
adaptive skeleton graphs perform equally well with neither holding a
decisive advantage.

\section{Discussion}

We have shown that the problem of finding shortest path and minimum
exposure path on a sensor network can be solved approximately with low
communication cost using skeleton graphs.  Our experiments confirm
that both the uniform and adaptive skeleton graphs provide close to
optimal paths with very low communication overhead.  Although, in the
asymptotic limit of large networks, adaptive skeleton graph is more
scalable, this is not a real issue for realistically sized networks.
Moreover as we have noted in the end of sec. \ref{sec-embedding},
there exists simple load balancing schemes for uniform skeleton graph.
Thus from the perspective of a practical implementation, the uniform
skeleton graph is superior to the adaptive graph in terms of its
simplicity and  load distribution.


\bibliographystyle{IEEEtran} 
\bibliography{biblio}

\end{document}